\documentclass[aps,prb,preprint,superscriptaddress,citeautoscript]{revtex4}
\usepackage[latin1]{inputenc}
\usepackage{dcolumn}
\usepackage{graphicx}
\usepackage{fixmath}
\usepackage[colorlinks=true,linkcolor=black,%
citecolor=black,urlcolor=black]{hyperref}

\usepackage{amssymb}
\usepackage{amsmath}
\usepackage{amsthm}
\usepackage{bm}
\usepackage{multirow}
\usepackage{psfrag}
\usepackage{mathrsfs}
\usepackage{footmisc}

\graphicspath{{./pics/}}

\newcommand{\itap}{Institut f\"ur Theoretische und Angewandte Physik (ITAP), %
  Universit\"at Stuttgart, Pfaffenwaldring 57, 70550 Stuttgart, Germany}

\newcommand{\ifmq}{Institut f\"ur Funktionelle Materialien und
  Quantentechnologien (IFMQ), %
  Universit\"at Stuttgart, Pfaffenwaldring 57, 70550 Stuttgart, Germany}

\newcolumntype{d}{D{.}{.}{3}}

\begin{document}
\title[Effectivity of Double Pulses]{Simulation of Laser Ablation
  in Aluminum: The Effectivity of Double Pulses}

\date{\today}

\author{Johannes Roth}
\affiliation{\itap}
\altaffiliation[New Address: ]{\ifmq}
\email{johannes@itap.physik.uni-stuttgart.de}
\author{Armin Krau\ss}
\author{Jan Lotze}
\author{Hans-Rainer Trebin}

\begin{abstract}
Lasers are becoming a more and more important tool in cutting and shaping
materials. Improving precision and effectivity is an ongoing demand in science
and industry. One possibility are double pulses. Here we study laser ablation
of aluminum by the two-temperature model. There the laser is modeled as a
source in a continuum heat conduction equation for the electrons, whose
temperature then is transferred to a molecular dynamics particle model by an
electron-phonon coupling term. The melting and ablation effectivity is
investigated depending on the relative intensity and the time delay between 
two Gaussian shaped laser pulses. It turns out that at least for aluminum the
optimal pulse shapes are standard Gaussian pulses. For double pulses with
delay times up to 200 ps we find a behavior as observed in experiment: the
ablation depth decreases beyond a delay of 10 ps even if one does not account
for the weakening at the second pulse due to laser-plasma interaction.
\end{abstract}

\maketitle

\section{Introduction}

Improving precision and effectivity of laser ablation is an ongoing demand in
industry and science. One possibility is given by double (or multiple) pulses
which have been studied intensively in experiment and numerical simulations. 

\subsection{Application of lasers}

Drilling holes with femtosecond laser pulses is assumed to deliver higher
quality than for example ps or ns-pulses \cite{semerok}. The drawback is that
for a single 
hole tens of thousands of pulses are required. With typical repetition rates 
in the kHz-regime this leads to pulse intervals in the range of
microseconds. From a molecular dynamics simulations point of view two such
subsequent pulses can be considered as independent since the sample cools down
completely between two pulses and the ablation plume vanishes. The only
difference is that the second pulse is applied to a damaged sample.  

The situation changes if the repetition rate increases and the time interval
between the pulses shortens. Studies of two subsequent femtosecond pulses show
that especially between about 10 ps and 10 ns the ablation depth is strongly
reduced \cite{semerok,axente,mildner}. Several models exist which try to
explain this effect \cite{semerok,axente,pova}. To get more insight from
molecular dynamics simulations combined with a two-temperature model explained
below we will
study double femtosecond pulses at delay times up to 200 ps. The absorption
of the laser pulses may be increased by an ablation plume which cannot be
modeled directly in molecular dynamics simulations. To take this effect into
account we have studied double pulses of different height. 

Shorter delay times between two pulses have to be produced in experiment by
special methods.
First a single Gaussian pulse is generated and sent through a pulse shaper
where it is modified with a liquid crystal modulator to produce
a sequence of femtosecond laser pulses with increasing or decreasing
intensity \cite{weiner,englert}. The sequences can then be interfered by
birefringent optical elements to generate Gaussian pulses with any desired
time delay. 

If the delay time is very short, the pulses become overlapping and can be used
to model non-standard pulse shapes which are not Gaussian in time. In covalent
materials it is obvious that the pulse shape plays an important role, since if
the front of the pulse is steep, it will produce many charge carriers which
can transport the laser energy into the bulk while if the front of the pulse
is slowly increasing it will only heat the sample locally. The question we
will address is how strong this effect will be in a metal where free electrons
are already present. For this goal we have studied overlapping Gaussian pulses
where the first pulse is lower or higher than the second pulse, thus called
increasing and decreasing pulses, respectively. 

\subsection{Studies of pulse sequences}\label{reason}

Semerok and Dutouquet \cite{semerok} find in experiments that the crater
depths generated by double pulses are twice as deep as from a single pulse for
delay times less than 1 ps, while they are only as deep as from a single pulse
for delays larger than 10 ps with a transient regime in between. They
attribute this result to reheating of the plume. 
Axente et al. \cite{axente} confirm this result by studying the brightness of
the ablation plume. 
Papadopoulou et al. \cite{papa} obtained opposite results for the ablation of
oxides: up to 1 ps the effectivity rises sharply and stays constant up to 10
ps. Longer time delays have not been considered.
Povarnitsyn et al. \cite{pova} studied double pulses theoretically with a
continuum model, and could reproduce the results obtained by Semerok and
others.  
They attribute the reduced effectivity of double pulses with delays longer
than 10 ps between the two pulses to the fact
that every pulse generates a shock wave followed by a rarefaction wave. The
ablation depth will be reduced if the shock wave of the second pulse
interferes with the rarefaction wave of the first pulse. If no rarefaction
wave is present, they attribute the reduced ablation depth to the pressure
exerted by heating the ablated material\cite{pova2}. 
Very recently Mildner et al. \cite{mildner} have studied laser ablation in
aluminum by laser-induced breakdown spectroscopy (LIBS) spectroscopy with time
delayed double pulse from 
femtoseconds to nanoseconds. They observe that four regimes exist: delay times
less than 1 ps (I), up to 10 ps (II), up to 100 ps (III) and above 100 ps
(IV). These regimes agree with those found by Semerok and Dutouquet
\cite{semerok} in experiments and by Povarnitsyn et al. \cite{pova} in theory. 

It is not possible to carry out simulations in a range far above ablation
threshold as it is done in experiment. 
The reason is that the ablation depth is of the same order of magnitude as our
sample size and it would take many nanoseconds to achieve a steady state of
ablation and melting. Together with the number of simulations for the required
data points this is out of reach for our computational capacity.
Therefore we have studied two cases: one set of simulations where we have
determined the ablation threshold for pulse shapes and double pulses. We could
also derive information about the melting depth, the electron and lattice
temperature and the distribution of kinetic energy.
In another set of simulations we applied fixed laser energies per area
(i.e. fixed laser fluence) above the ablation threshold as in experiment. Here
we are limited to a regime far below the experimental range, nonetheless a more
direct comparison to experiment is feasible. 

The paper is organized as follows. We first present the two-temperature model
(TTM) coupled to molecular dynamics simulations (MD). 
After some simulation details we describe the pulse types studied in this work. 
Following are results and a discussion of the two cases: double pulse ablation
close to the ablation threshold, and double pulse ablation at fixed fluence
above the threshold. The paper is concluded with a summary.

\section{Two temperature model coupled to molecular dynamics 
simulations}   

\subsection{Equations}

In femtosecond laser ablation the laser radiation acts directly on the free
electrons of the metal, exciting them to a non-equilibrium state. If the
electrons are equilibrated fast enough it is possible to define a temperature
for them. This is the base of the two-temperature model (TTM) of 
Anisimov et al. \cite{Anisimov1974} where separate temperatures for the
electrons and the lattice are introduced. The equation for the electron
temperature is:
\begin{equation}
  C_{e}(T_{e}) \frac{\partial T_{e}}{\partial t} = \nabla [K_{e} \nabla
  T_{e}] - 
  \kappa (T_{e} - T_{i}) + S(\textbf{x},t). \label{eq:TTM1}
\end{equation}
$T_{e}$ and $T_{i}$ are the electron and ion temperatures, $C_{e}$ is the heat
capacity, $K_{e}$ the heat 
conductivity, $\kappa$ the electron-phonon coupling constant and
$S(\textbf{x},t)$ the external laser field. To work on an atomistic scale the
equation for the lattice temperature is replaced by the Newtonian equation of
motion of the atoms extended by a coupling term which acts on the velocities
(for more details see \cite{ivanov}~):
\begin{equation}
   m_{j} \frac{d^{2}\textbf{x}_{j}}{dt^{2}} = - \nabla_{\small \textbf{x}_{j}}
  U(\lbrace \textbf{x}_{k} \rbrace) - \frac{\kappa}{C_{l}} \frac{(T_{i}
  -T_{e})}{T_{i}} m_{j} \textbf{v}^{T}_{j}.
\end{equation}
$U$ is the interaction potential, $\textbf{x}_{k}$ are the coordinates,
$m_{j}$ the atom masses, $\textbf{v}^{T}_{j}$ the thermal velocities
of the atoms. The coupling parameter $\kappa$ has to be translated to
atomistic observables by $C_{l}$, the atomistic heat capacity which is
computed directly from the simulations. $T_{i}$ and $C_{l}$ are determined
locally by averaging over slabs containing several hundreds of atoms. 

Instead of the electronic heat capacity $C_{e}$ which is a linear function of
temperature, for metals at moderate temperatures the heat capacity coefficient
$\gamma=C_{e}/T$ is introduced which is constant over a broad temperature range.

\subsection{Simulation details, boundary conditions}

All simulations have been carried out with IMD, the {\bf I}TAP
{\bf M}olecular {\bf D}ynamics simulation package
\cite{imd1,imd2}. The simulations were run on our local computing nodes with
up to 12 CPU, and on the Cray XE6 Hermit of the Stuttgart computing center
using 640 compute cores. The total number of runs was 423, each lasting about
one day.

The material studied in all simulations was aluminum. The interaction of the
atoms has been
modeled by the EAM potential of Ercolessi and Adams \cite{Ercolessi1994}. 
For the electronic parameters of Al we set the coefficient for the specific
heat of the electrons $C_{e}=\gamma T$ to $\gamma=135$ J/(m$^{3}$$K^{2})$
\cite{26}, the electron-phonon coupling constant $\kappa=2.45 \cdot 10^{17}$
J/(m$^{3}$Ks) \cite{26}, and the electronic heat conductivity $K_{e}=235$
J/(sKm) \cite{33}.
 
The sample size for the simulations at the ablation threshold was a box with
241.92$\times$4.84$\times$4.84 nm$^{3}$ containing about 350'000 atoms. At the
beginning the sample was equilibrated at 305 K.
The sample size for the simulations with high fluence was a box with
364.46$\times$12.15$\times$12.15 nm$^{3}$ containing 3'240'000 atoms.  At
the beginning the sample was equilibrated at 293 K. 

The boundary conditions are periodic in the transverse direction and open along
the direction of the laser beam. The samples for short pulse delay time are
chosen long enough that during observation time no interaction occurs
between the ablation procedure and the waves which are emitted towards
the rear end of the sample, are reflected back and return to
the surface. This limits the simulation time for a 1000 nm sample to about 300
ps since the speed of sound in solid Al is about 6400 m/s. The time is reduced
to about 200 ps if the sample melts due to a lower speed of sound of about
4700 m/s.

For long delay times prohibitively long samples would be
required. Fortunately, the pressure waves can be removed by applying a 
ramp at the rear end of the sample with the 
following properties: starting at 70\% of the length of the probe a damping
force is added to the equations of motion which increased quadratically
towards the end of the probe and removes the kinetic energy of the wave (for
details see Ref.\ \cite{imdlas}).

\subsection{Determination of melting and ablation depth.}
\label{abldepth}

Melting and ablation depths are determined from $x$-$t$ histograms of
the density (see Fig.\ \ref{meltdepth}). We monitor the motion of the
detached layer. If its distance to the bulk is constantly
increasing we note that ablation has set in. This procedure may be somewhat
imprecise if the samples are very large and thus the required simulation time
to obtain a steady state is very long. Furthermore, the formation and
collapse of bubbles below threshold may interfere with the true ablation
threshold.
Preuss et al.\ \cite{preuss104} determine the ablation threshold $F_{th}$ and
the inverse absorption length $\alpha^{-1}$ by fitting the observed ablation
depth $d_{abl}$ at an applied fluence $F$ to the equation
\begin{equation}
d_{abl}=\frac{1}{\alpha}\ln \left(\frac{F}{F_{th}}\right).
\end{equation}
This method could give more accurate results, but it would also require many
time-consuming simulation runs to determine a single ablation
threshold. Therefore we stick here to our simpler visual method.
The melting and ablation depths are given with respect to the
unexpanded sample. Therefore, and since the ablated material is molten, the
melting depths include the ablation depths (see also Fig.\ \ref{meltdepth}).  

The literature values for the single pulse ablation threshold of Al vary in a
broad range.
In experiments Guo et al.\ \cite{guo52} have observed first damage of Al
samples occurring at a fluence (energy per area) of 340 J/m$^{2}$
already, while Vorobyev \cite{vorobyev130} determine an ablation threshold of
$F_{th}=530$ J/m$^{2}$ in air and 580 J/m$^{2}$ in vacuum. Le~Harzic et al.\
\cite{leharzic73} on the other hand find $F_{th}$ as high as 1200 J/m$^{2}$. 
In numerical studies Anisimov et al.\ \cite{anisimov6} computed a value of
$F_{th}=700$ J/m$^{2}$. 

With the method of Preuss et al.\ \cite{preuss104} in previous work
\cite{SonntagAPA} we had obtained $F_{th}=(858 \pm 170)$ J/m$^{2}$, for the
Ercolessi and Adams potentials. This 
threshold seems to be much lower than the values which we find in this
article. Several reasons are responsible for this fact: 
first of all,  in Ref.\ \cite{SonntagAPA} there are large error bars in the
determination of the ablation threshold which would justify an ablation
threshold up to 1200 to 1400 J/m$^{2}$ instead of the reported $F_{th}=858$
J/m$^{2}$. 
Second, the ablation threshold of $F_{th}=858$ J/m$^{2}$ is an extrapolation
to an ablation depth $d_{abl}=0$ nm, whereas in simulations we always need to
observe a finite ablation depth $d_{abl}$ to detect the process. Thus our
method will overestimate the ablation threshold.
Third, the ablation threshold is difficult to determine precisely since it
depends crucially on the initial conditions. To get good statistics many
simulations with different initial conditions would have to be carried
out. It is known that the threshold obtained in simulation depends on the size
of the sample\cite{zhigileiprep}.
The transverse dimension of the sample plays an important role since
it determines if the ablated material still forms a layer or if it breaks up
into droplets. Larger transverse dimensions would be desirable, but this would
lead to huge runtime demands.

\section{Simulation Procedure and Pulse types}

\subsection{Properties of two-Pulse Sequences}
 
The cross sections of the simulated samples are small compared to the typical
half width of a laser spot studied in experiment. Therefore we used a constant
illumination of our samples with no transverse beam dependence.
The laser beam intensity distribution in Eq.\
(\ref{eq:TTM1}) is $S(x,t)=(1-R)\cdot \mu \exp{(-\mu x)} \cdot\sigma_{E} \cdot
g(t)$. The reflectivity $R$ and the inverse absorption depth $\mu^{-1}$ are
set to the typical values of aluminum: $R=0.87$ and $\mu^{-1}=8$
nm for a wave length of 800 nm \cite{SonntagAPA,33}. 
$g(t)$ is the temporal shape of the laser pulse. 
The fluence of the laser beam is given by $F=\sigma_{E}/(1-R) \cdot
\int_{0}^{\infty} g(t) {\rm d} t$.
The original pulse has Gaussian shape with width $\sigma_{t}=0.4$ ps:
\[
g_{0}(t)=\frac{1}{\sqrt{2\pi}\sigma_{t}}\exp{\left(-\frac{1}{2}
\frac{(t-t_{0})^{2}}{\sigma^{2}_{t}}\right)}.
\]

\begin{figure}
\centering
\includegraphics[width=3.7cm,angle=270]{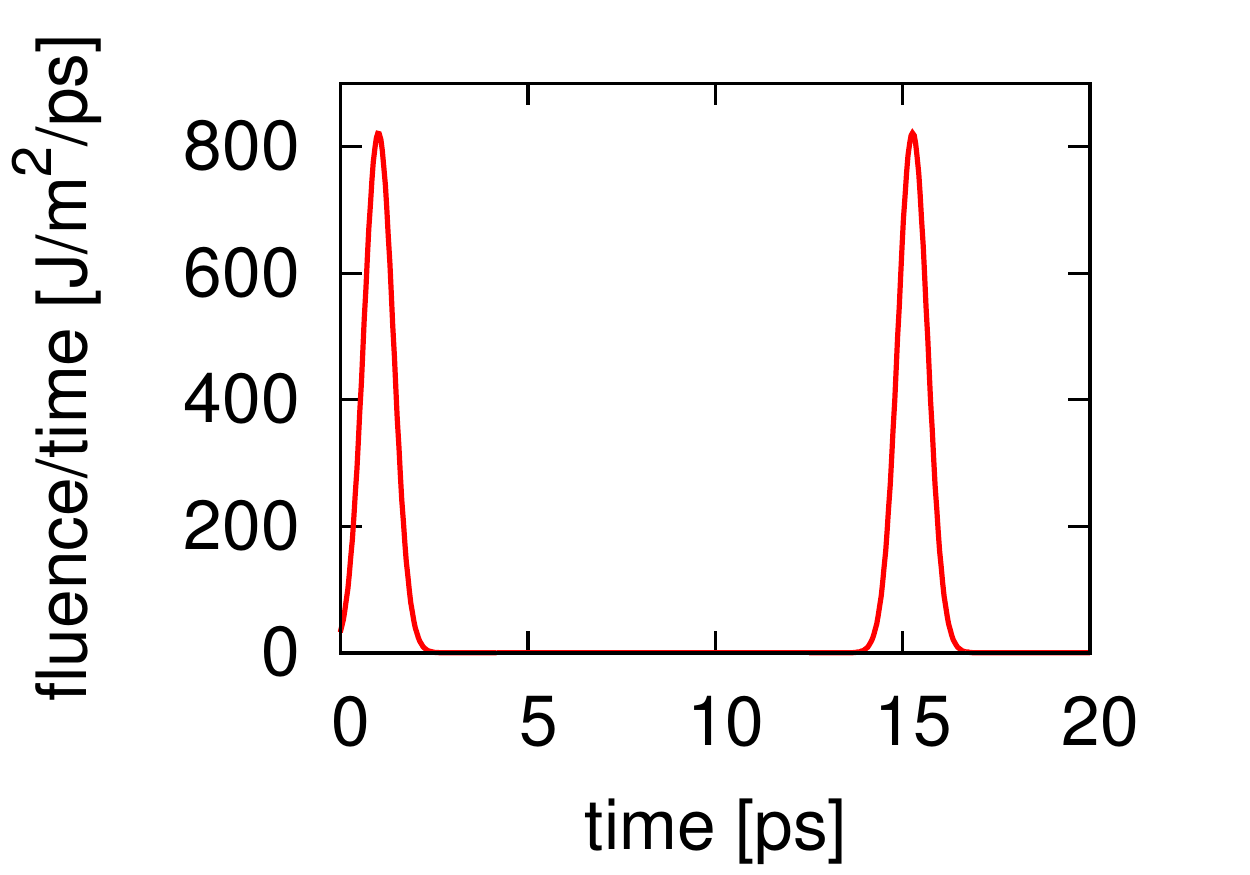}
\includegraphics[width=3.7cm,angle=270]{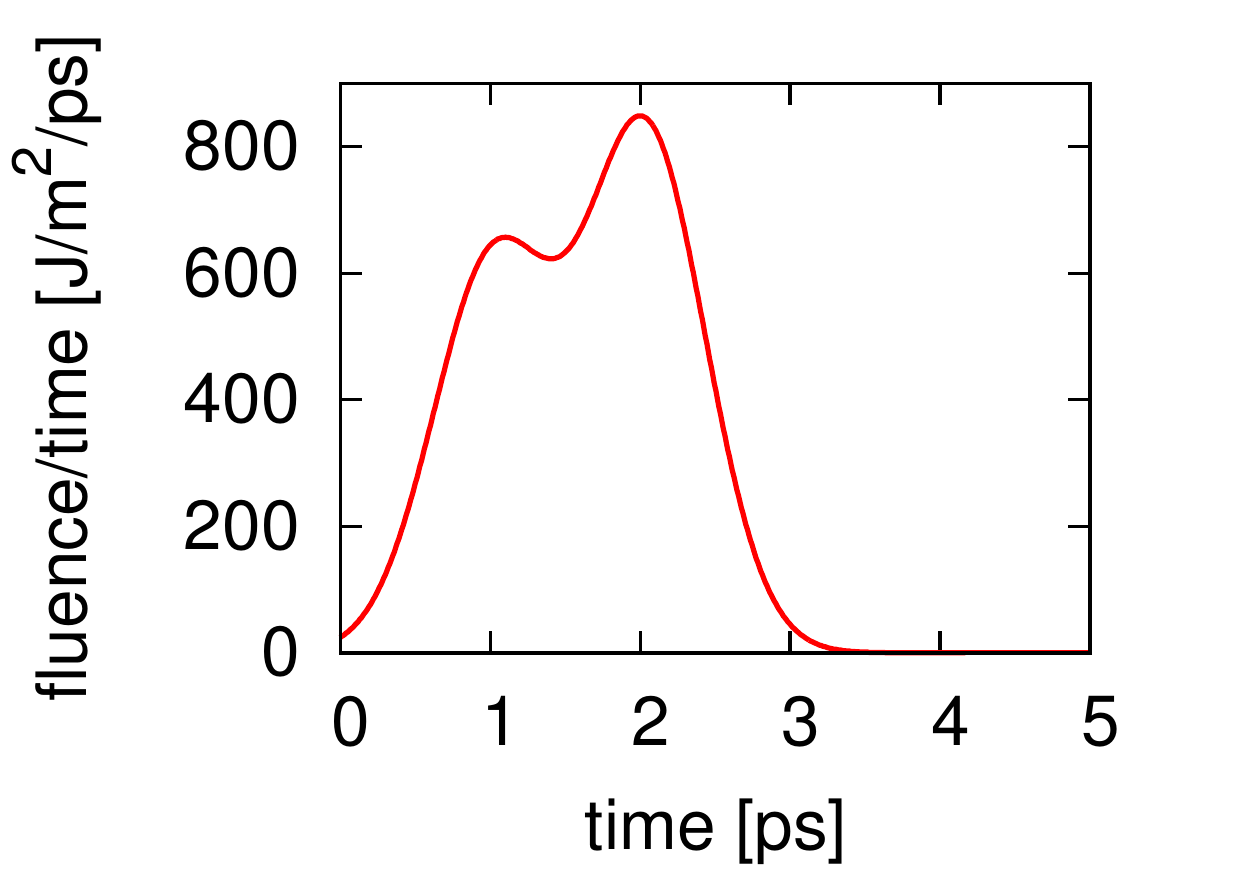}
\includegraphics[width=3.7cm,angle=270]{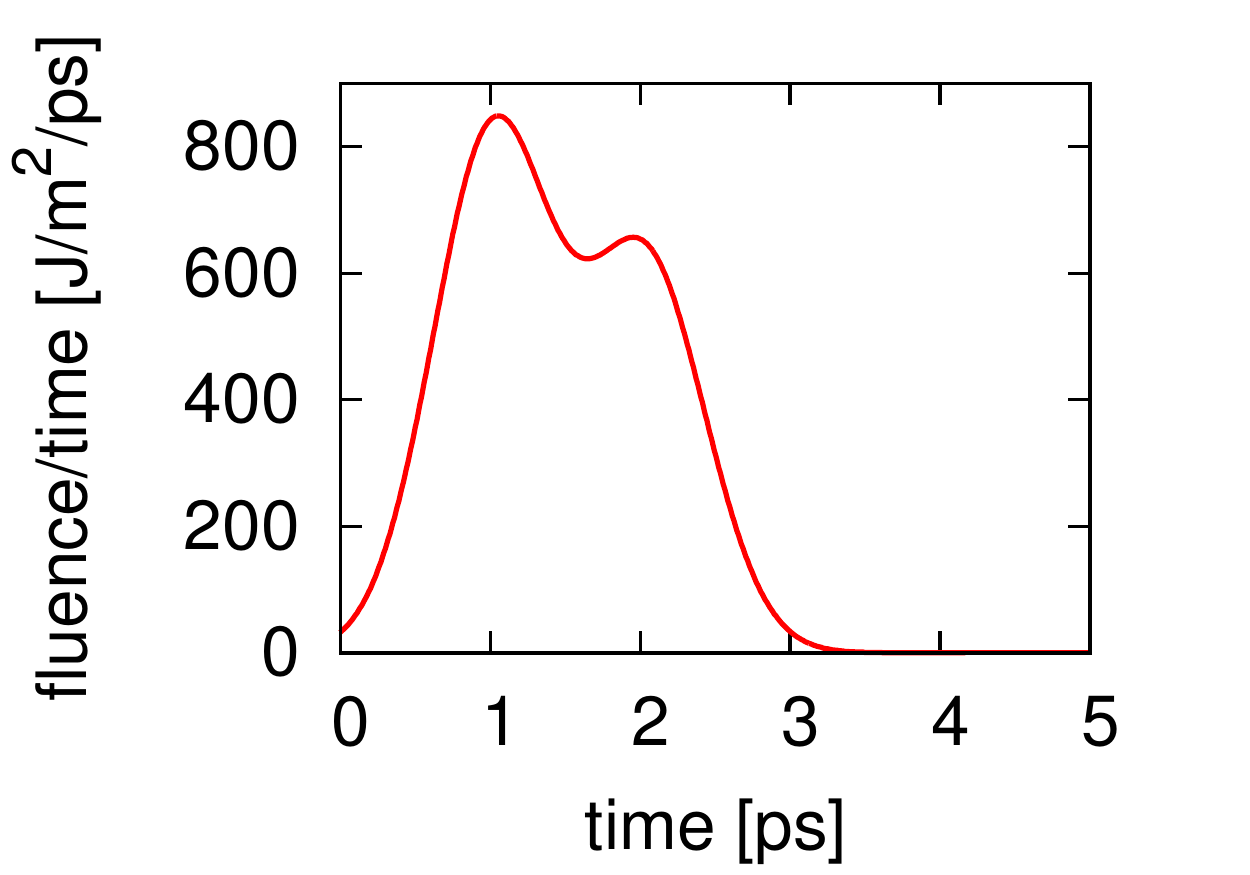}
  \caption{Pulse shapes. From Left to right: double pulse at a time interval of
    $\Delta$t = 15ps, overlapping pulses at a time interval of $\Delta$t =
    2ps, increasing with relative height 3 and 4, and decreasing with relative
    height 4 and 3.\label{pulseshape}}  
\end{figure}

The general formula for the shape of the simulated pulses is
\begin{eqnarray}
g_{i}(t)&=& \frac{1}{\sqrt{2\pi}\sigma_{t}} \left[a_{i}
\exp{\left(-\frac{1}{2}\frac{(t-t_{0})^{2}}{\sigma^{2}_{t}}\right)+} \right.
\nonumber \\ 
 &+& \left. b_{i}\exp{\left(-\frac{1}{2}\frac{(t-c_{i} \cdot
      t_{0})^{2}}{\sigma^{2}_{t}}\right)} \right], \nonumber
\end{eqnarray}
with amplitudes $a_{i}$ and $b_{i}$ and time interval $c_{i}$.
Three scenarios were studied with the following $g_{i}$:
\begin{enumerate} 
\item[$i=1$] are called \textbf{dou}ble pulses; here $a_{1}=b_{1}=1$ and
  $c_{1}$ is variable (see Fig.\ \ref{pulseshape} left for 
$c_{1}=15$), or $a_{1}=1$, $b_{1}$ variable and $c_{1}=15$,   
\item[$i=2$] is called an \textbf{inc}reasing pulse with $a_{2}=3/4$,
  $b_{2}=1$ and 
  $c_{2}=2$ (Fig.\ \ref{pulseshape} center),  
\item[$i=3$] is called a \textbf{dec}reasing pulse with $a_{3}=1$,
  $b_{3}=3/4$ and 
  $c_{3}=3/4$, (Fig.\ \ref{pulseshape} right),  
\end{enumerate}

The technical parameter $t_{0}=1.018$ ps is used to shift the pulse such that
it hits the sample surface after the simulation starts. 

Sequences of simulations were carried out by increasing the 
parameter $\sigma_{E}$ at intervals of $\Delta\sigma_{E}=8$
J/m$^{2}$ until ablation occurs. Thus the data given for the ablation
thresholds are accurate only within about $\Delta\sigma_{E}/2$. 

\section{Results}

First we report results from simulations where the ablation threshold was
determined for a standard Gaussian pulse and the non-Gaussian increasing and
decreasing pulse shapes. This section includes the detailed analysis of one
double pulse with the parameters $a_{1}=b_{1}=1$, and $c_{1}=15$. Double
pulses with variable $c_{1}$ are treated later in Sec.\ \ref{const}.

\subsection{Simulations close to the ablation threshold}\label{thresh}

\begin{figure}
\centering
\includegraphics[width=8cm]{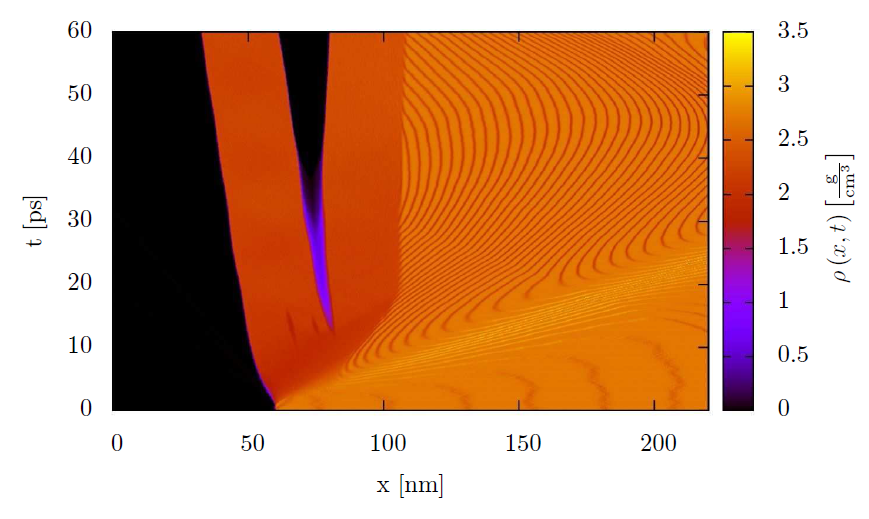}
\includegraphics[width=8.3cm]{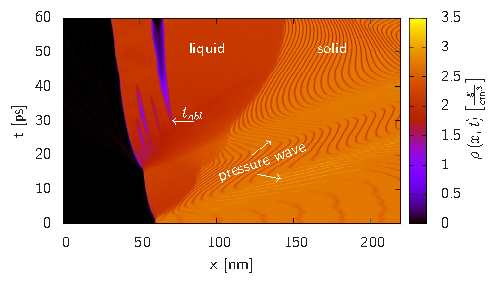}

\includegraphics[width=8cm]{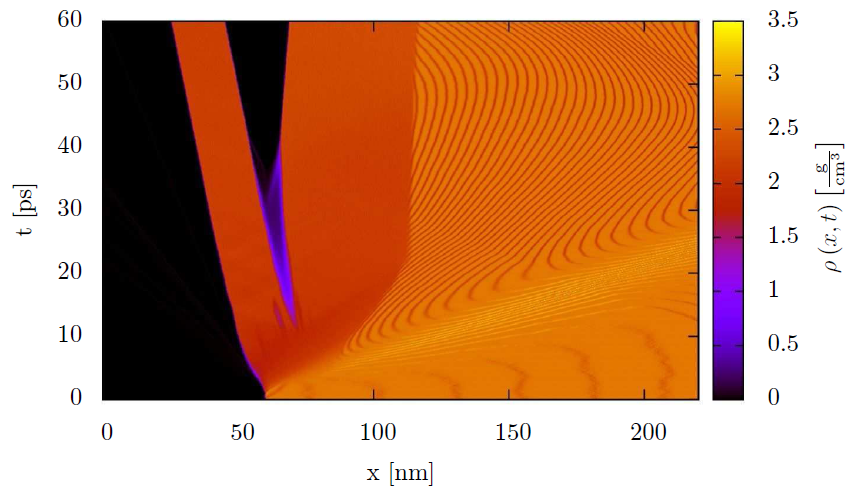}
\includegraphics[width=8cm]{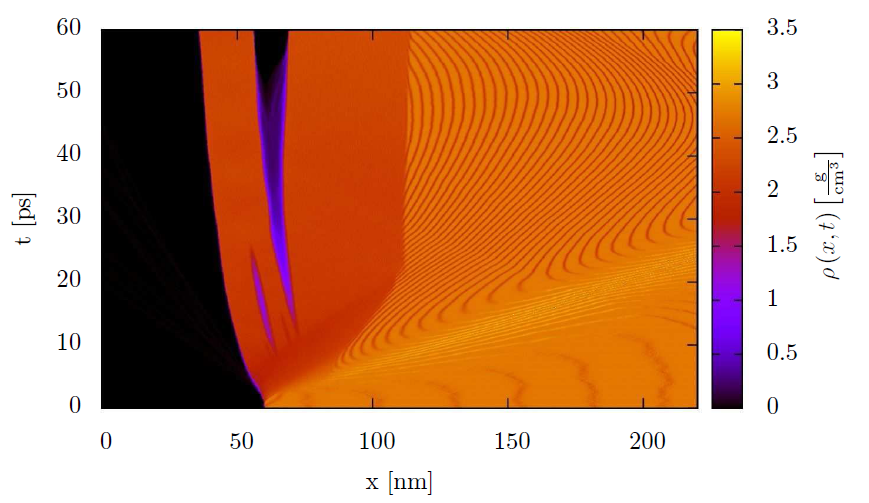}
  \caption{Melting depths, from left to right and top to bottom: single
    pulse, double pulse, increasing pulse, decreasing
    pulse. The pictures show the temporal evolution of the local density
    histograms, averaged about the cross section of the
    samples. The laser beam is applied from the left side at time
    $t_{0}=$1.018 ps \label{meltdepth}} 
\end{figure}

\textbf{Melting and ablation behavior}. 
In general, the samples start to melt, then ablation sets in and the samples
continue to melt down to a constant depth. Modifications occur for non-Gaussian 
pulses and the double pulse: the ablated layers are not generated in one step
but split up. For the double pulse there is a clear distinction between the
action of the first and second pulse (see Fig.\ \ref{meltdepth}).  
The quantitative results are summarized in Tab.~\ref{tabparabs}. 
Ablation typically occurs about 12 to 15 ps after the laser beam has hit the
sample.  Thus there is no interaction with the shock wave emitted into the
bulk since the wave returns to the surface much later.
\begin{table}
\centering
\begin{tabular}{|cc|c|c|c|c|}
\hline
\multicolumn{2}{|c|}{Pulse type}& single & double & increasing & decreasing\\ 
\hline
$F_{th}$ & [J/m$^{2}$]& 1260 & 1648 & 1439 & 1367\\
\hline
$t_{abl}$ & [ps] & 12.6 & 29.8 & 11.9 & 12.2\\
\hline
$d_{abl}$ &[nm] & 21.2 & 26.0& 20.0& 20.3\\
\hline
$d_{melt}$ & [nm] & 45.3 & 79.6&54.0& 51.0\\
\hline
 
\end{tabular}
\caption{Summary of the sample behavior for single and double pulses and for
  increasing and decreasing pulse shapes as shown in Fig. \ref{pulseshape}. 
$F_{th}$ is the fluence to achieve ablation, $t_{abl}$ is the time when the
  ablation process starts, $d_{abl}$ and $d_{melt}$ are the depths of the
  ablated and the molten layers, respectively.} 
\label{tabparabs}
\end{table}

We note that the double pulse requires a fluence of 1648 J/m$^{2}$ to achieve
ablation which is less than twice the fluence of 1260 J/m$^2$ for ablation by
a single pulse. It means 
that the fluence of a pulse can be reduced if a second pulse follows, but with
respect to the total applied fluence this is not advantageous. The fluence 
goes down to less than 1461 J/m$^{2}$ if the time delay between the pulses is
reduced to 10 or 8 ps. The ablation threshold fluence further decreases
continuously to the value of a single pulse if the time delay goes to $\Delta
t=0$ ps.

The effective applied laser fluence is reduced by the interaction of the laser
beam with the ablation plume which is equivalent to the increase of the
reflectivity of the sample. This effect 
cannot be simulated directly. To mimic it, we have lowered the height of the
second pulse in four steps down to half the height of the first pulse. The
total applied fluence necessary for ablation rises slightly to 1710 J/m$^{2}$
which means that the first pulse nearly reaches the fluence of a single
pulse (855 and 1260 J/m$^{2}$, respectively). There seems to be no significant
and systematic change in the ablation 
depth and the starting time of ablation while the melting depth increases due
to the higher fluence.  

\textbf{Temperatures}. The electron and phonon temperatures of the samples
have been determined as function of time at a position $x=60$ nm deep in the
bulk.

\begin{figure}
\centering
\includegraphics[width=8cm]{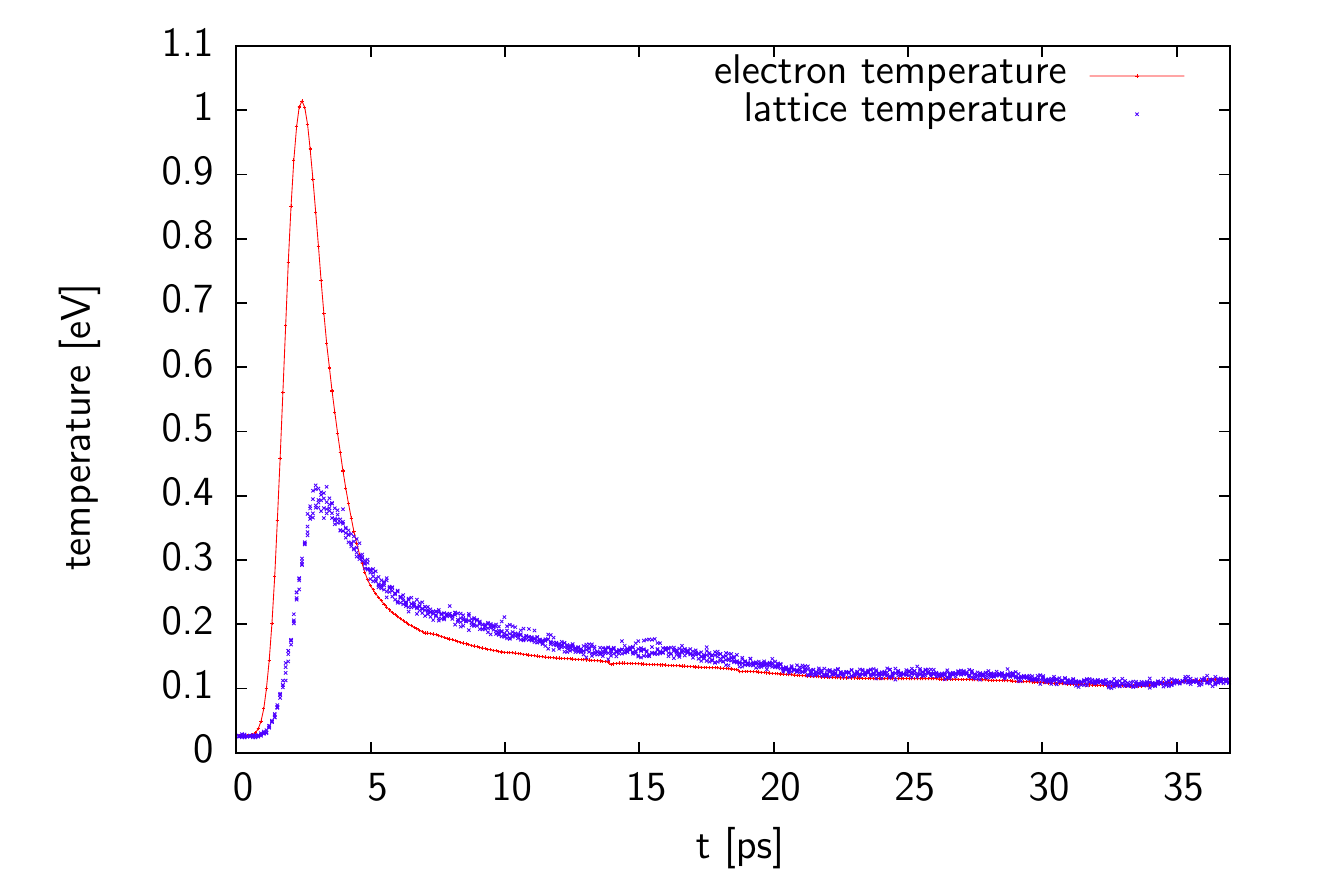}
\includegraphics[width=8cm]{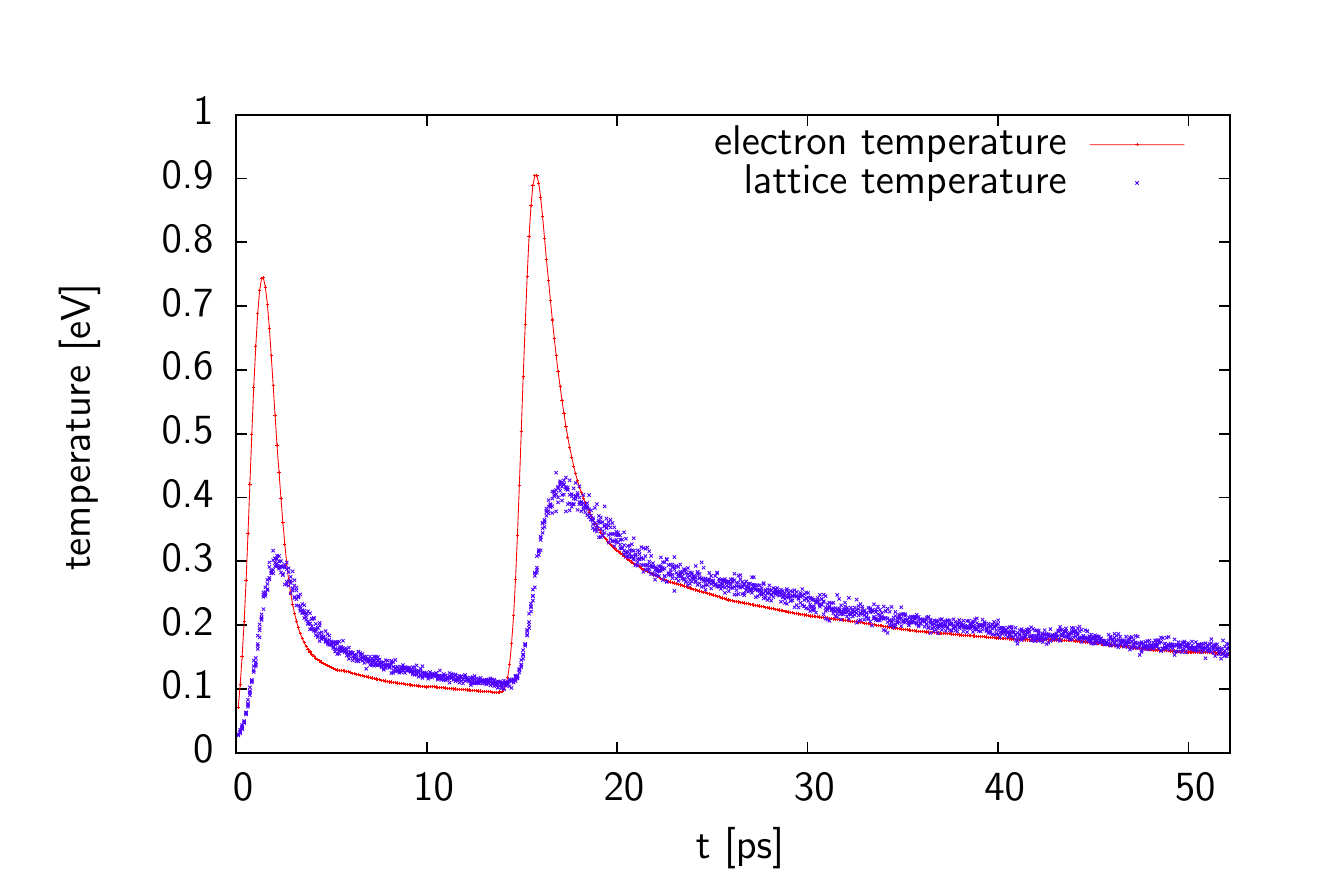}

\includegraphics[width=8cm]{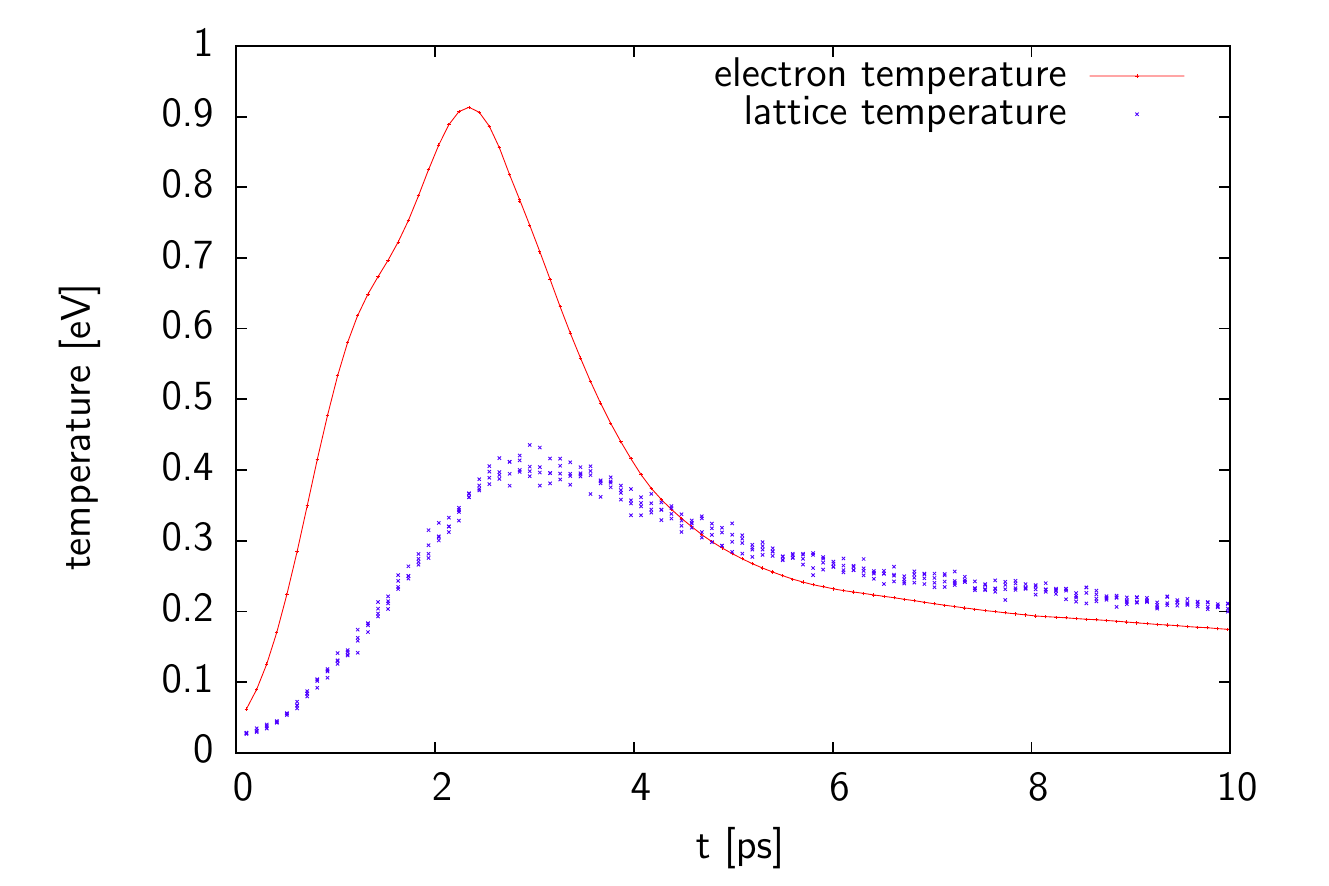}
\includegraphics[width=8cm]{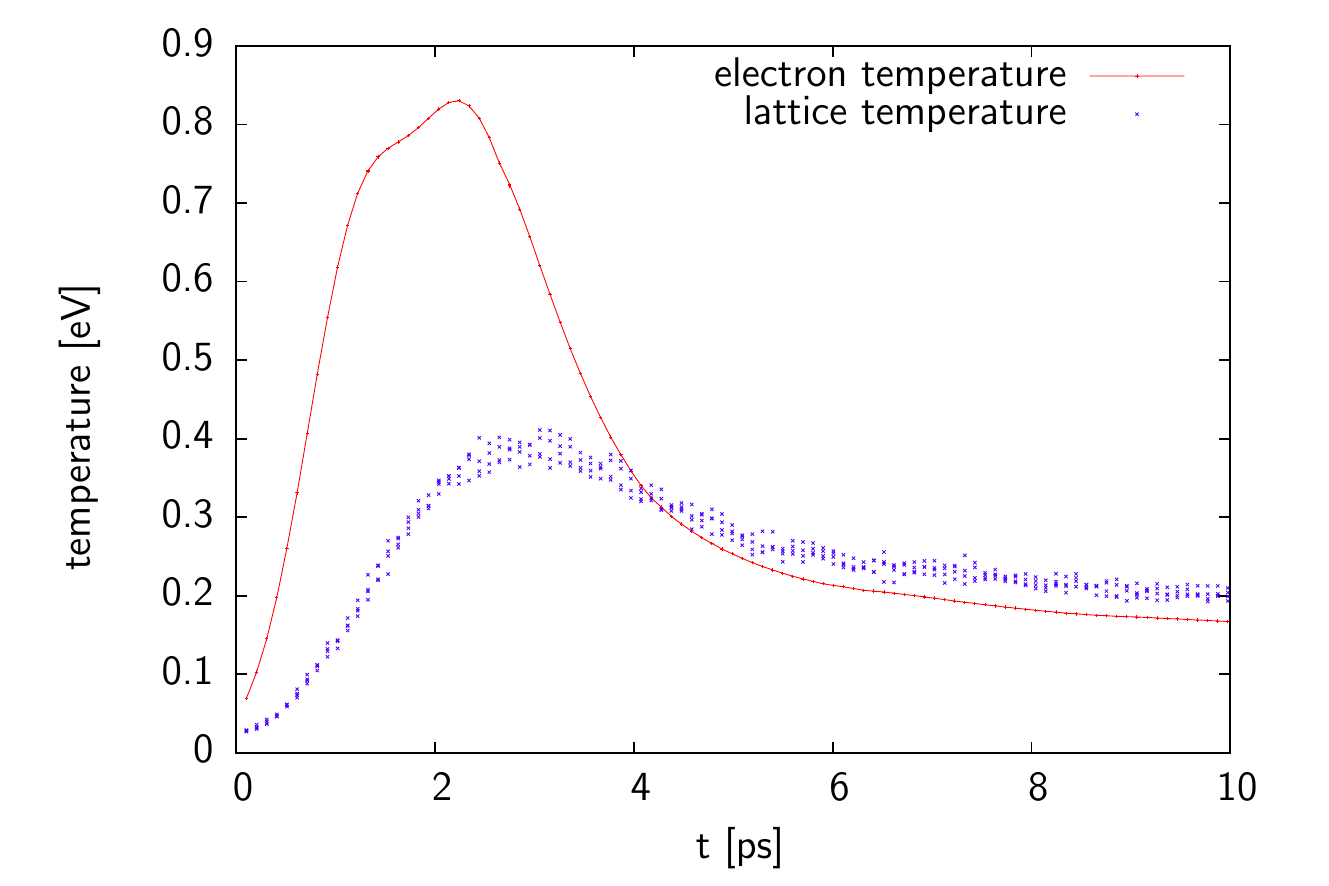}
  \caption{Lattice and electron temperature at a sample depth of $x$ = 60
    nm. From left to right and top to bottom: single pulse, double pulse,
    increasing pulse, decreasing pulse.\label{temp}}   
\end{figure}

For the double pulse the electron and lattice temperature have nearly reached
equilibrium when the second pulse hits the sample (see Fig.\
\ref{temp}). Since this temperature is about the same as the temperature reached
for a single pulse at 35 ps, we can consider the two pulses as separate
pulses. The temperature at the end of the double pulse simulation at 50 ps is
significant higher than for the single pulse due to the larger fluence applied.

For the increasing pulse (Fig.\ \ref{temp}) the evolution of electron and
lattice temperature follows the signature of the pulse shape, although the
minimum of the pulse is smeared out. The overall behavior is rather similar to
the single pulse, the electron energy is slightly lower and the maxima are
broader.
 
For the decreasing pulse the slope of the electron temperature is as sharp
as for a single pulse, but then it continues to rise further at a slower
pace. This behavior is caused by the onset of the second pulse maximum at that
time. The lattice temperature behaves similar to the one of the increasing
pulse but it is broadened further. The second lower maximum leads to a further
decrease in 
the peak of the electron energy and in a stronger broadening of the maximum. 

In conclusion we find that the non-Gaussian overlapping maxima behave similar to
a Gaussian with the same fluence but a broader with $\sigma_{t}$.

\begin{figure}
\centering
\includegraphics[width=8cm]{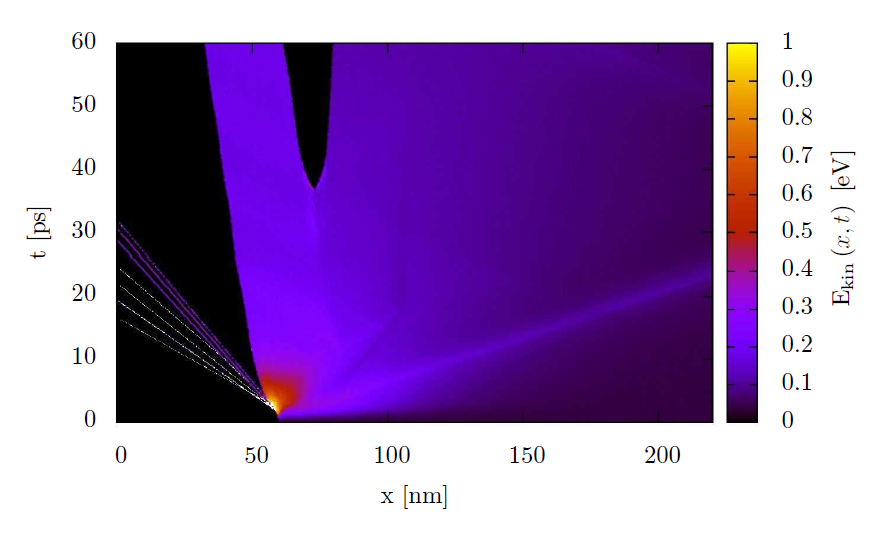}
\includegraphics[width=8.3cm]{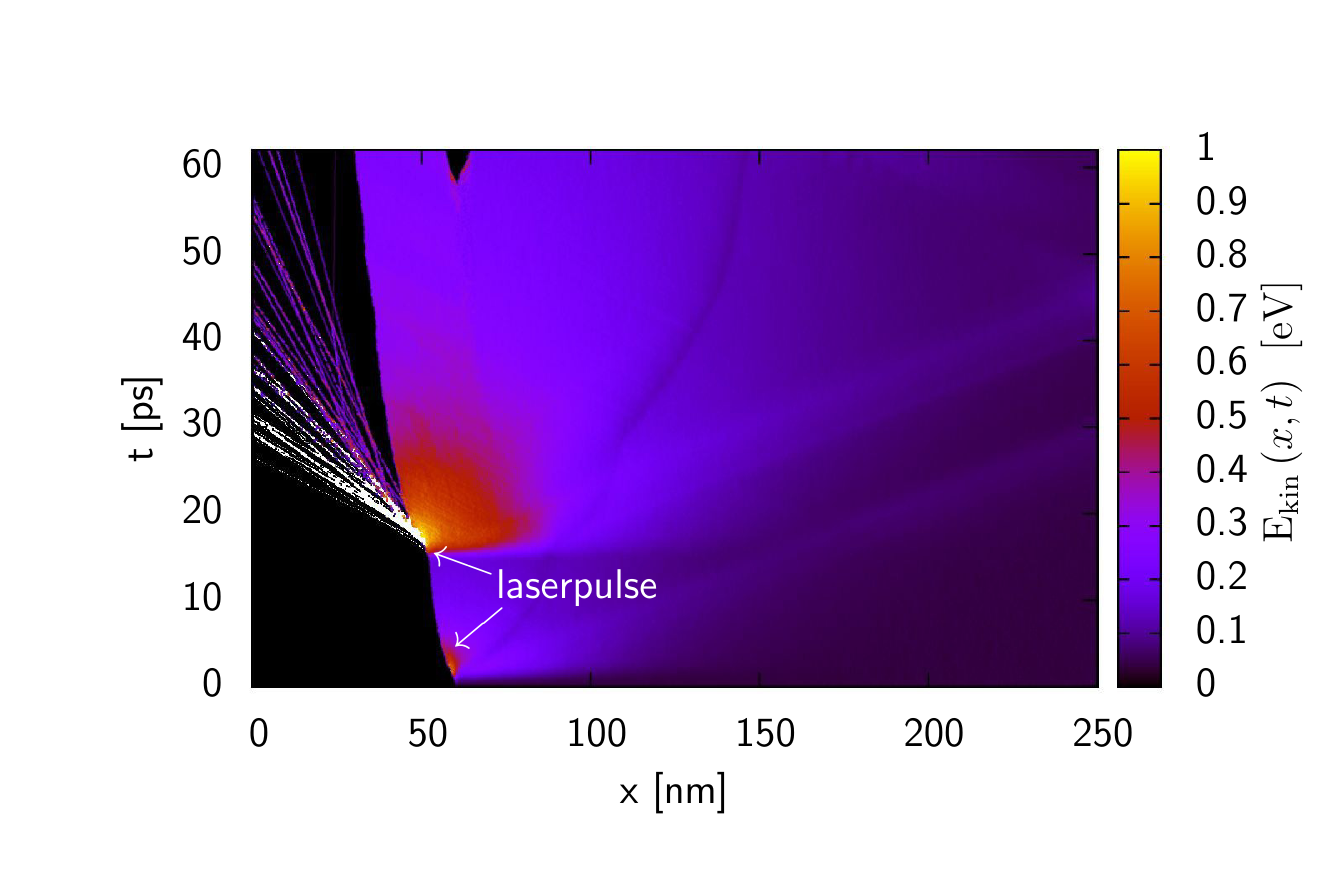}

\includegraphics[width=8cm]{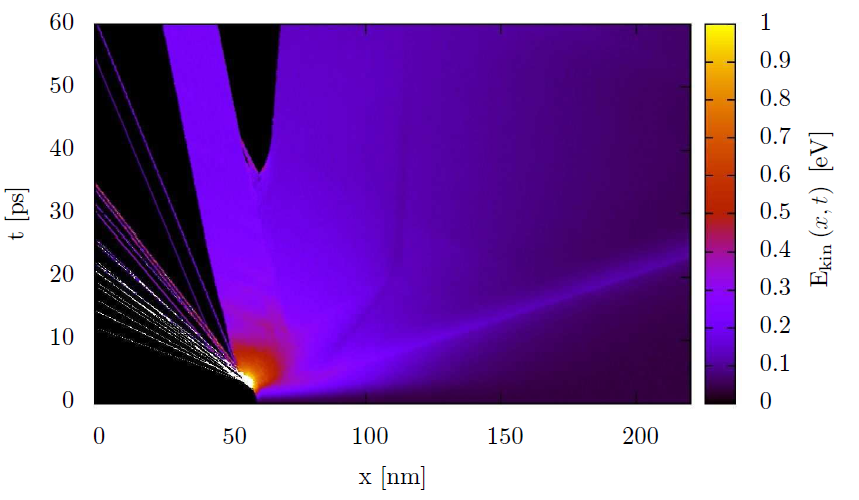}
\includegraphics[width=8cm]{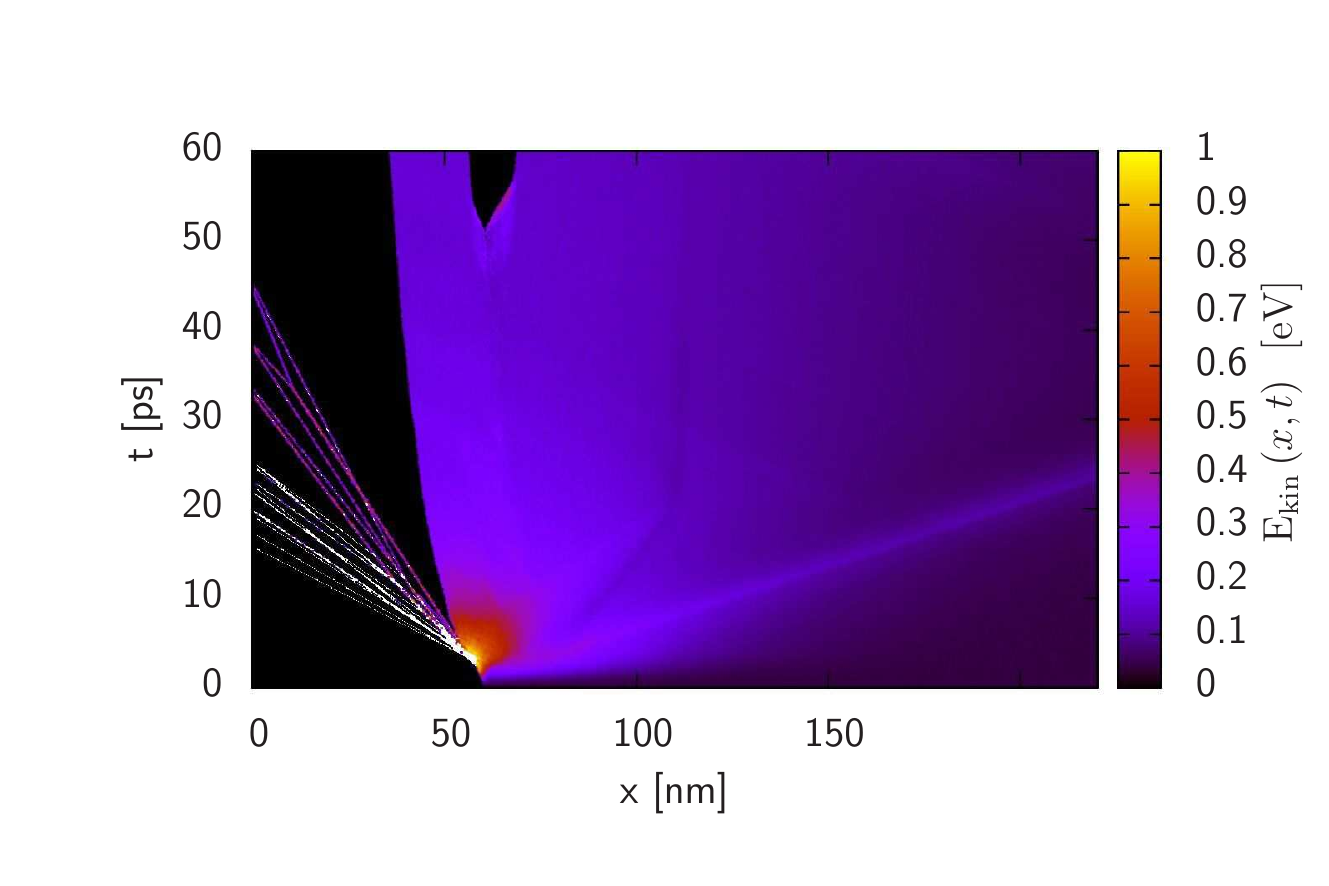}
  \caption{Kinetic energy, from left to right and top to
    bottom: single pulse, double pulse, increasing pulse, decreasing
    pulse. The pictures show the temporal evolution of the local kinetic
    energy, averaged about the cross section of the
    samples. \label{kineten}}
\end{figure}

\textbf{Kinetic energies}. Fig.\ \ref{kineten} shows the distribution of the
local kinetic energy \emph{vs.}~time. The laser pulse is applied at the hot
spot. It is clearly visible that jets of single atoms are emitted immediately
after excitation and prior to full layer ablation. This indicates that the
samples are indeed damaged already far below the ablation threshold.

\textbf{Pressure waves}.
Each laser pulse generates a sharp shock wave moving into the bulk of the sample
which is followed by a broad rarefaction wave. Povarnitsyn et al. \cite{papa}
find that the rarefaction wave of the second pulse 
is attenuated with respect to the rarefaction wave of the first pulse.
Furthermore, the depth of the rarefaction wave of the first pulse stays 
constant whereas the rarefaction wave of the second pulse gets weaker with
increasing delay time.

The intensity of the rarefaction wave in our simulations is 3.6 GPa for the
single pulse. The height of the shock wave is about 15.6 GPa at a sample
depth of 100 nm and a fluence of 810 J/m$^2$. For the double pulses at a delay
of 15 ps we find at the same depth that the rarefaction wave of the second
pulse is reduced to 2.3 GPa but is \emph{stronger} than the rarefaction
wave of the first pulse at 1.2 GPa. Unfortunately, we do not know the strength
of the rarefaction wave of the second pulse for longer delay times, but for
shorter delays it is similar to the 10 ps case. The total fluence of the double
pulses was 1647 J/m$^2$ and the peak intensity of the shock wave of the
first pulse was 9.5 GPa.
Interestingly, we observe that the shock wave of the second pulse is strongly
attenuated by the rarefaction wave of the first pulse; from 7.1 GPa at 15 ps
delay to 4.6 at 10 ps delay. Furthermore, an intermediate very sharp shock
wave is generated by interference of the shock waves of the first and second
pulse with peak heightd 6.5 GPa for 10 ps delay. 

\subsection{Simulations at fixed fluence}\label{const}

\textbf{Single pulses and general remarks}:
Ablation at fixed fluence and variable time delay between the two
pulses has been studied with the larger sample of about 3 million
atoms. 

Full ablation for single pulses has been observed at 2095 J/m$^{2}$.
Large bubbles start to grow already at 1848 J/m$^{2}$ but collapse after
100--150 ps. Thus the ablation threshold should be somewhere between these two
fluences. 
The determination of the ablation depth is problematic since it shows
strong random fluctuations and no systematic increase for higher fluences as
would be expected. 
Improving statistics would require many simulations with
different initial conditions. For single pulses an average ablation depth of 27
$\pm$ 5 nm was obtained in the range from 2000 to 3700 J/m$^{2}$.  

The initial goal was to simulate fixed fluences at variable time delay far
above ablation threshold and time delays as large as possible. For reasons
given in the introduction (Sec.\ \ref{reason}) this goal could not be achieved
fully. In experiments pulses have been used with fluences that are 20 to 60
times the ablation threshold\cite{mildner}. In simulations a total fluence
above 3700 J/m$^{2}$ leads to a complete destruction of the sample.
Therefore we could study only double pulses at fluences of 1848 and 2218
J/m$^{2}$.

Possible delay intervals between pulses are also shorter than in experiment. We
have studied delay intervals up to 200 ps, with good statistics
up to 10 ps. Longer delay intervals lead to very long simulation times even if
the damping ramp is applied since the molten material slows down the
simulation speed additionally due to the increased mobility of the atoms in
the melt as compared to atoms in a solid.

Thus the simulations are still close to the ablation threshold, but now we can
scan are large set of intervals systematically.

\begin{figure}
\centering
\includegraphics[width=8cm]{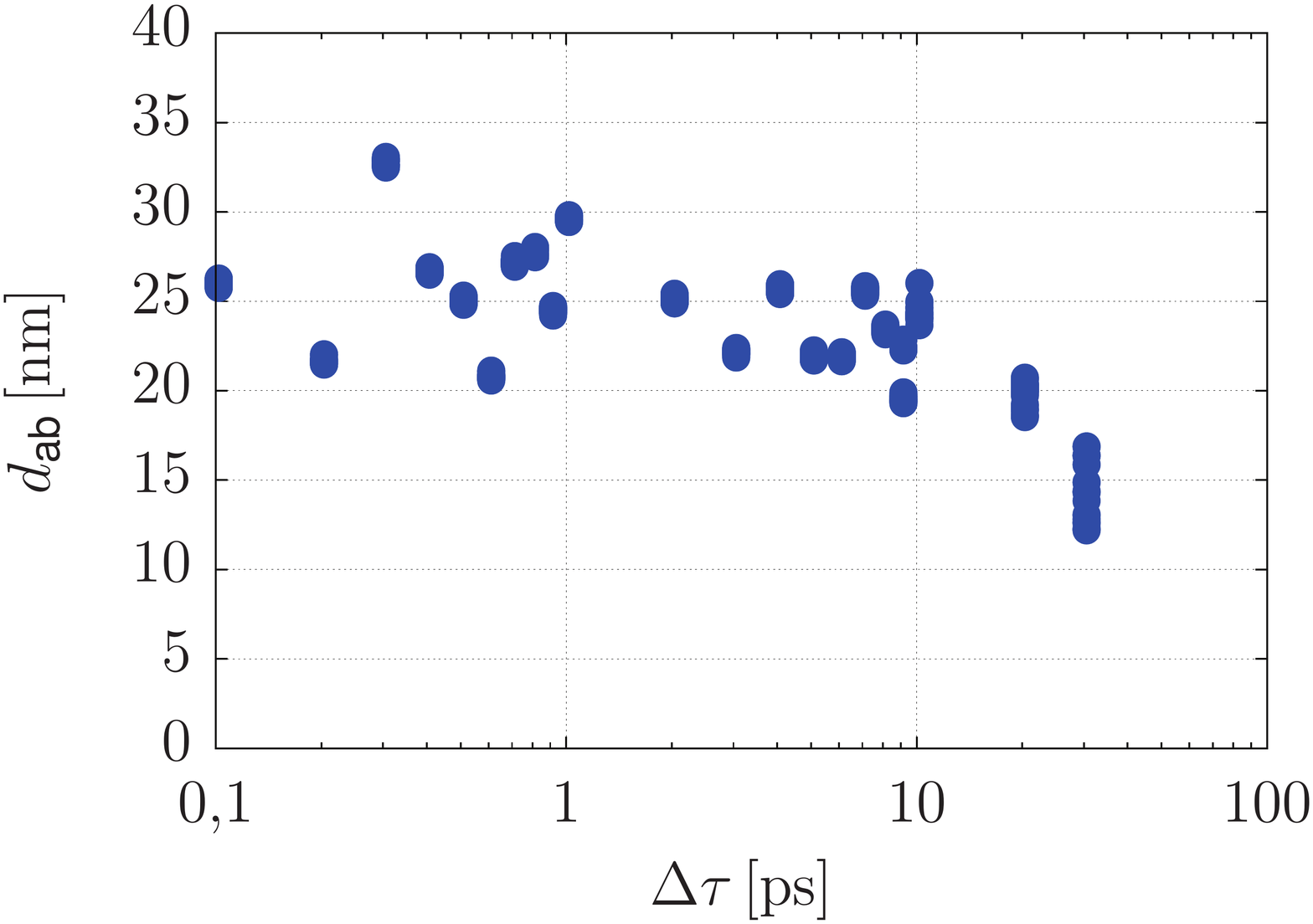}
\includegraphics[width=8cm]{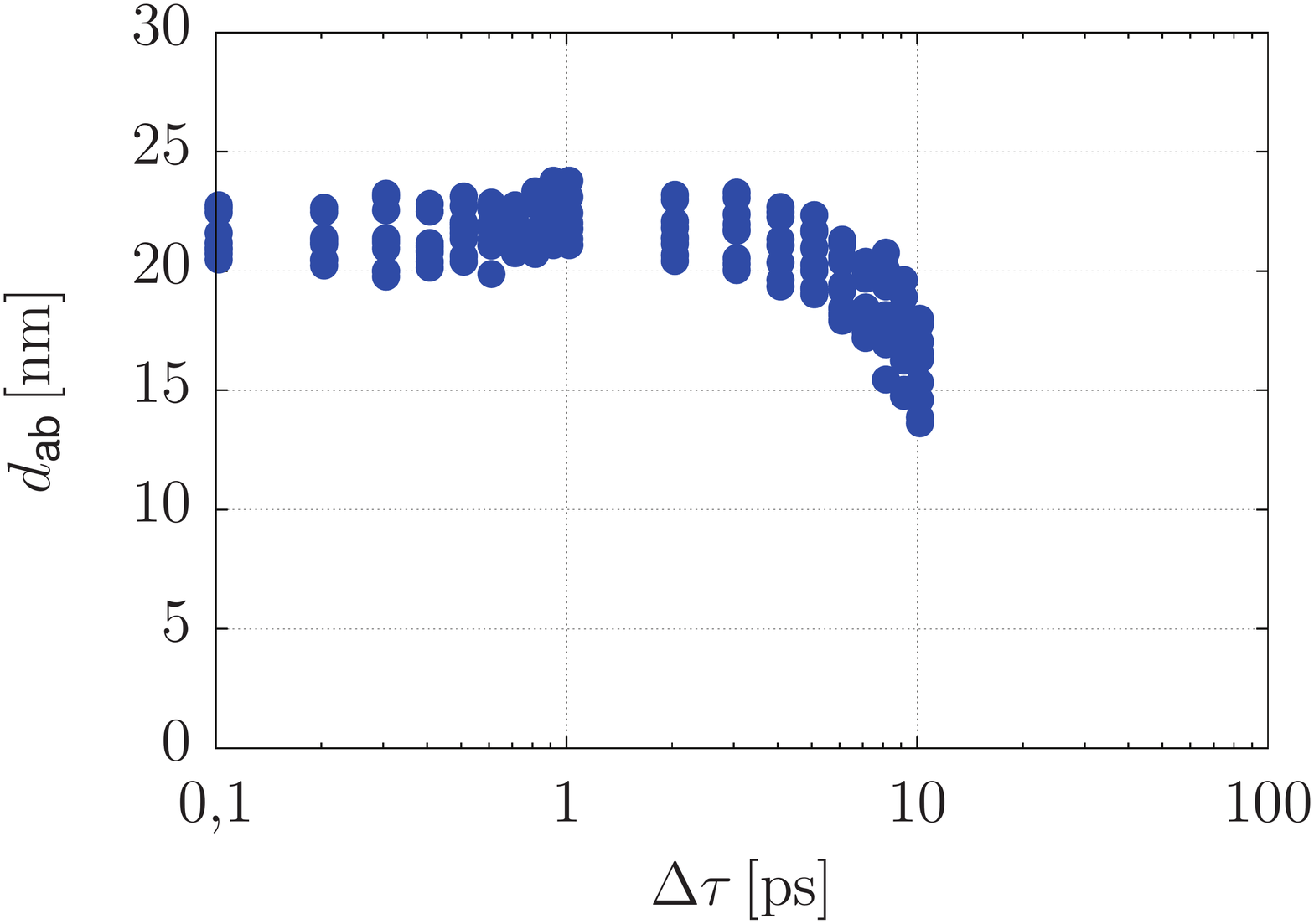}
  \caption{Ablation depths {\it vs.} time delay for double pulses. Left: at
    $F=1848$ J/m$^{2}$, right at $F=2218$ J/m$^{2}$. \label{depth}}
\end{figure}

\textbf{Double pulses}:
Both cases (1848 and 2218 J/m$^{2}$) behave quite
similar (see Fig.\ \ref{depth}). For the lower fluence a plateau is observed
at an ablation depth of $d_{abl} = 25\pm2$ nm up to a pulse delay of $\Delta t
= 10$ ps, followed by a decreasing ablation depth up to $\Delta t = 30$
ps. For the higher fluence the plateau is slightly lower  and ranges only up
to about 5 ps. The decrease is then observed up to $\Delta t = 10$ ps.
The ablation threshold appears to converge to a second plateau at about
$d_{abl} = 10\pm3$ nm for longer time delays in the range between $\Delta t =
20$ ps and 60 ps, but for this range the statistics is rather weak. The
behavior of the ablation depth for delay times up to 200 ps is rather
inconclusive, but it certainly does not increase again to the value of short
time delays. 

\begin{figure}
\centering
\includegraphics[width=8cm]{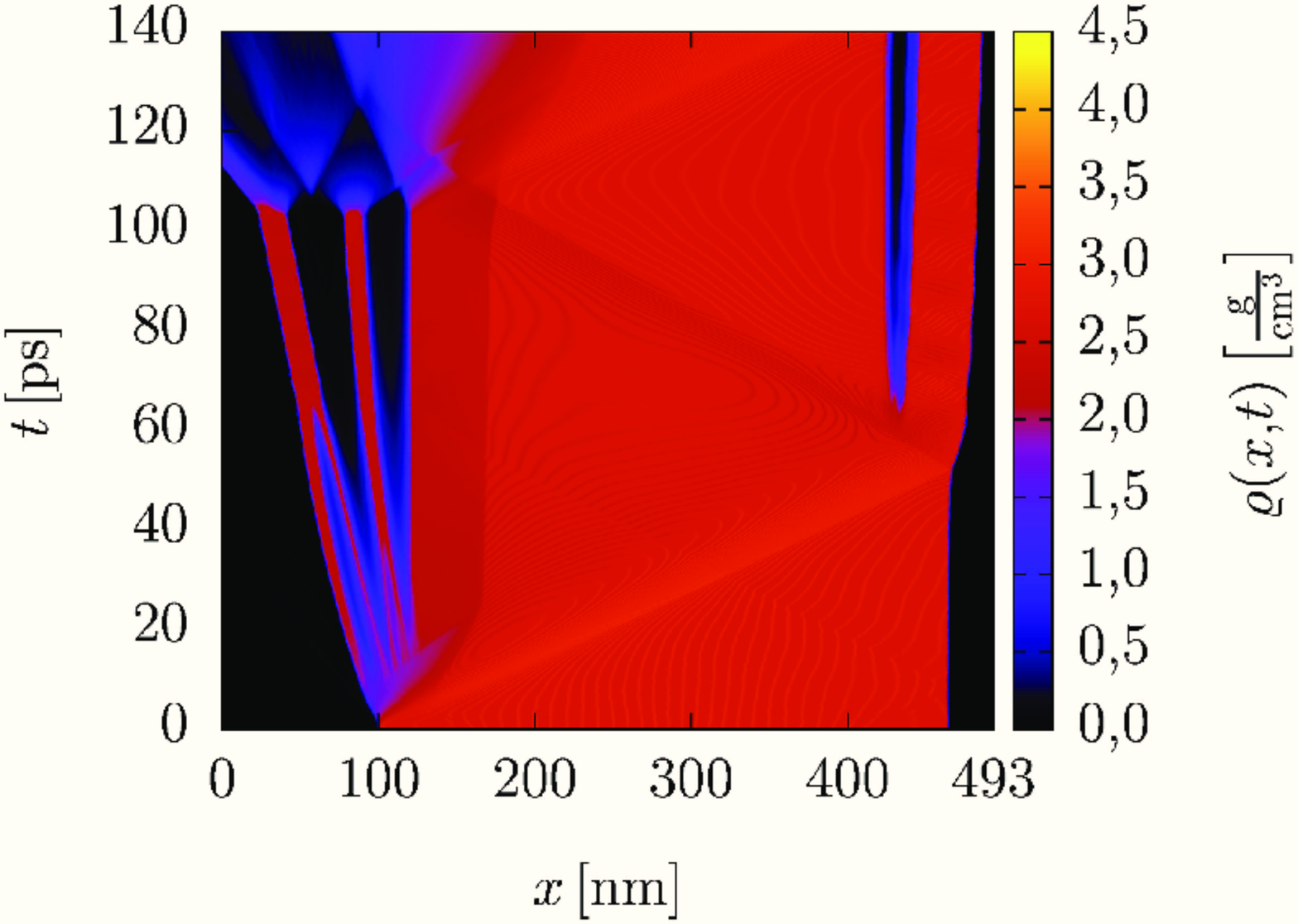}
\includegraphics[width=8cm]{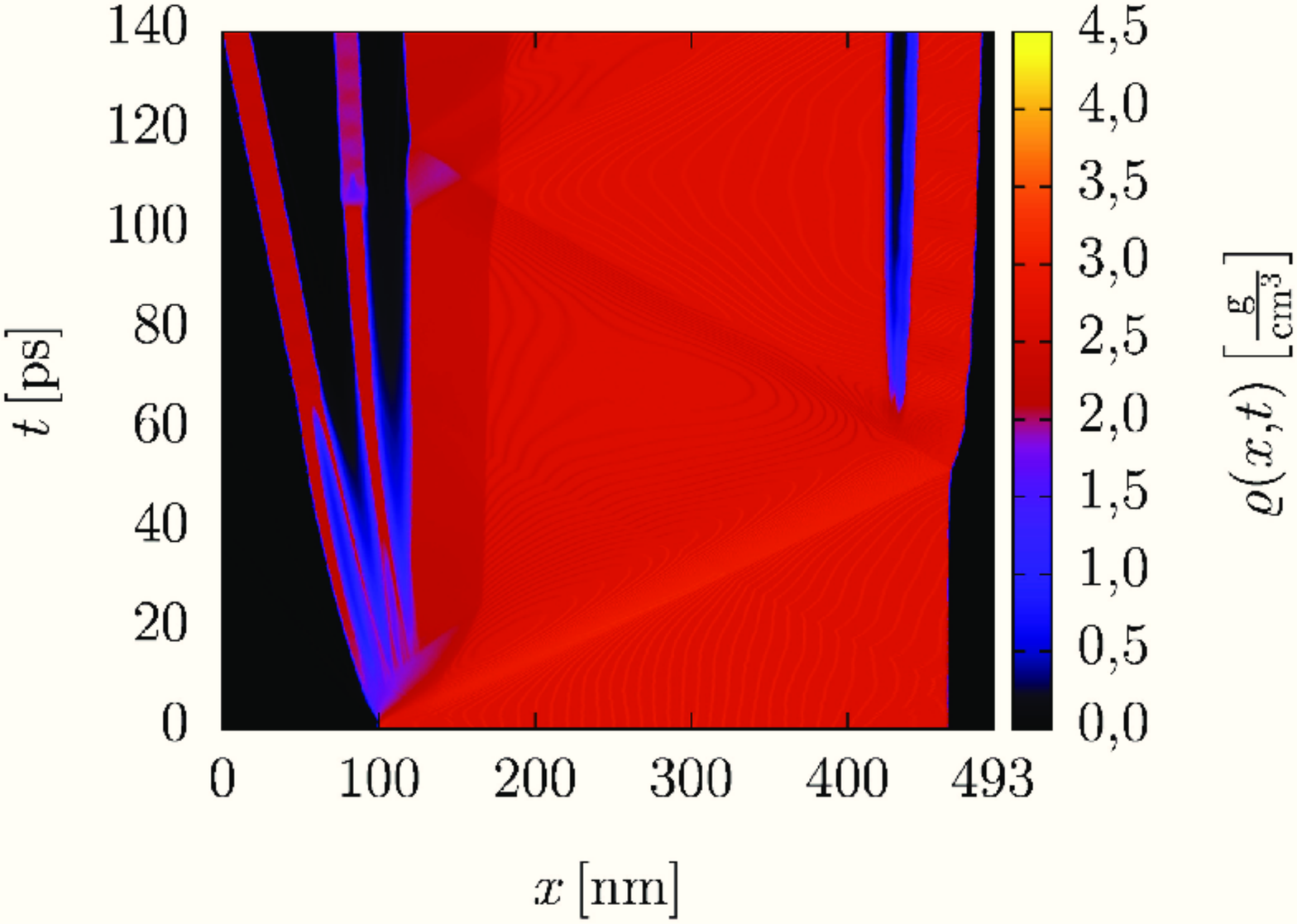}
  \caption{Density histogram of the ablation process of a sample with free
    rear surface. Left: at the upper left hand corner the vaporization of the
    plume is observed if the full laser intensity is applied above the new
    surface. Right: result if the second pulse is 
    applied directly to the surface and the interaction with the ablated
    material is taken into account by an attenuation factor. \label{cloud}} 
\end{figure}

\textbf{Pulse attenuation}: We are not (yet) able to model the
full interaction of the laser beam with the ablation plume including the
plasma. In Fig.\ \ref{cloud} it is clearly seen how the second pulse
interacts with the ablated material by vaporizing it. This scenario is not
realistic since the beam has to be applied \emph{above} the surface of
the sample, and the intensity is reduced globally. 
A more realistic approach is the following: 
we can determine a scattering cross section $\sigma=\mu \cdot m/\rho$
with the absorption length $\mu$, and atom mass $m$ and the density $\rho$ 
for aluminum if we assume for simplicity that the laser beam interacts with
the ablated material in the same way as with the bulk. Measuring the mass and
density of the ablated material it is thus possible to determine an attenuation
factor $f$ for the second pulse. This is certainly a crude model which can be
improved in future. In our simulations it results in a material and sample
dependent attenuation factor of $f\approx 0.0997$ which has been applied
directly after measuring the distribution of the ablated material during
simulation. The resulting difference can be seen in Fig.\ \ref{cloud} right.

\section{Summary and Discussion}

In the first part with simulations close to the ablation threshold we find that
two separate pulses require much more fluence than a single pulse to
achieve ablation while the final melting depth is much larger for two pulses
than for a single pulse. The reason is that the energy of the pulses adds up but
ablation is initiated by the first pulse only. 
If the height of the second pulse is between 75 and 87.5 \% of the first pulse
we find that the fluence has to rise up to about 1785 J/m$^{2}$ to achieve
ablation.
For the overlapping pulses the ablation threshold is about 15 \% higher than
twice that for the single pulse. This is due to the lower 
maximum and shows that it is better to concentrate the laser energy into a very
short pulse. The melting depths are larger for the overlapping pulses than for
the single pulse which reflects the fact that more energy has been added to
the system. This result is in good agreement with our previous study of the
melting depth of aluminum \cite{Sapporo} where we found that the melting depth,
which is more robust than the ablation depth, is largely independent of the
pulse duration and depends linearly on the applied fluence.

In the second study with fixed fluences and variable time delay
we find that within the range of fluences accessible by
simulations the ablation depth is more or less constant up to about 3700
J/m$^{2}$ with rather large errors. The reason for the uncertainty has been
explained in Sec.\ \ref{abldepth}. 

%
In general our results are in good qualitative agreement with experiments
\cite{semerok,mildner} and hydrodynamical modeling \cite{pova}. 
For the double pulses we observe a behavior similar to that obtained in
simulations for example by Povarnitsyn et al.\ \cite{pova} up to a time delay
of 500 ps. 
In accordance with them we find an attenuation of the rarefaction
wave of the second pulse for delay times starting at least at 15 ps, but in
contrast to \cite{papa} this wave is still deeper than the rarefaction wave of
the first pulse. Thus we could agree that the attenuation is responsible for the
reduction of ablation efficiency.  
On the other hand we observe an additional effect, namely that for short times
the shock waves of the first and second pulse interfere constructively, thus
enhancing ablation efficiency or at least keeping it on a high level.

Our results may also be compared to the work of Rosandi and
Urbassek \cite{Rosandi2010}. The pulses they study are half as wide as ours,
and they vary the time between the two pulses. Unfortunately they did not
report melting depths and ablation thresholds; thus a quantitative comparison is
not possible. They claim that the second pulse increases the formation of
voids and ablation if it returns during a pressure maximum while the opposite
the pulse acts during a pressure minimum. The modulation of the pressure occur
for pulse delays of 10 ps and 2.5 ps, but not for 4 ps. We do not find
evidence of such an influence in our simulations. 

The major shortcoming of the simulations which limits the comparability to
experiment is the limitation of the simulations to "low" fluences rather
close to the ablation threshold. Although there is some space to improve the
performance our simulation the only solution is to study much larger samples
and to carry out many more simulations to improve statistics.

Other experimental results for covalent materials show that the most effective
way to ablate material by multi-pulses is a decreasing sequence of pulses with
very short distance between them, almost overlapping. For increasing sequences
the small peaks at the beginning do not lead to melting or ablation and
thus this pulse shape is not as effective as decreasing pulses. The most
effective pulse shape, however, seems to be a single sharply rising pulse with
a slow decay \cite{Cola11}.

\section{Conclusion}

We have presented a study of a combined molecular dynamics simulation and
two-temperature model applied to femtosecond laser ablation.
We find that the specific pulse shape is rather insignificant for metals,
shorter pulses lead to higher electron temperatures and require less fluence
since the energy is more concentrated. 
Double pulses have no advantage over single pulses at delays up to hundreds of
picoseconds, and this is even more significant if the second pulse is reduced
through absorption.

\section{Acknowledgment}

This work has been supported by the German Research Foundation DFG as part of
the Collaborative Research Center SFB 716 in subproject B.5.

\end{document}